\documentclass[preprint,12pt]{elsarticle}
\usepackage{color}
\usepackage{graphicx}
\usepackage{subfig}
\usepackage[font=footnotesize,labelfont=bf]{caption}
\usepackage{amsmath,amsfonts,amssymb}
\usepackage{pdfsync}
\usepackage[para]{threeparttable}
\usepackage{hyperref}
\usepackage[version=3]{mhchem}
\biboptions{sort&compress}

\newcommand{\f}{\frac}

\graphicspath{{figs/}}
\newcommand*{\OrigAA}{}
\let\OrigAA\AA

\renewcommand*{\AA}{%
	{\fontfamily{ptm}%
		\selectfont%
		\OrigAA%
		\selectfont}%
}
\journal{Journal of Molecular Liquids}
\begin{document}
	\title{Global isomorphism approach: attractive Yukawa fluid, 2D case}
	\author[ak]{A.~Katts\corref{cor1}}
	\ead{a.katts@stud.onu.edu.ua}
	\author[vk]{V.~Kulinskii}
	\ead{kulinskij@onu.edu.ua}
	\cortext[cor1]{Corresponding author}
	\address[ak,vk]{Faculty of Mathematics, Physics and Information Technologies, Odesa National University, Dvoryanskaya 2, 65082 Odesa, Ukraine}
\begin{abstract}
In this paper we apply fluid-lattice gas global isomorphism approach to liquid-vapor equilibrium of Yukawa attractive fluid in the two dimensions. The construction of tangent to the binodal of the fluid in the low-temperature region is performed based on the Zeno-element. The dependence of Zeno-element  and Boyle parameters on interaction parameters is also studied. It is shown that the asymptotic behavior of the Zeno-element parameters can serve as a marker for liquid phase instability in this case. We also provide the relation between critical temperatures of bulk (3D) and monolayer (2D) fluid.
\end{abstract}
\begin{keyword}
	Zeno-line \sep 2D Fluids \sep law of corresponding states \sep Yukawa fluid \sep critical point \sep Ising model
\end{keyword}
\maketitle
\section{Introduction}\label{sec:intro}
Locating the phase equilibrium and its dependence on the properties of the interaction potential play an important role in the study of fluid systems, such as simple and associative liquids, micellar solutions, and fullerenes. One of the key aspects here is the stability region of the liquid phase and the shape of the gas-liquid binodal. It is known that the equilibrium liquid phase may only exist if the attraction is strong enough \cite{liq_fullersoliliq_nature1993,eos_yukawafluid_jcp1994,eos_liqvap_jcondmat1997,eos_ljmetastable_physa1999}. One can turn to computer simulation methods \cite{book_frenkelsimul} or try to solve the Ornstein-Zernike equation with help of different closure procedures  \cite{book_hansenmcdonald,liq_integraleq_physrep1996,eos_sarkisov_jcp2001}. Though it might be analytically difficult to perform for complex enough potentials. There is also well developed thermodynamic similarity concept but seemingly it can not be directly applied for finding the stability region of the liquid phase. It is rather tempting to search for some simple markers of instability based on the general properties of low-density states like virial coefficients \cite{eos_virial_prl2012}. The empirical rule Vliegenthart-Lekkerkerker  $B(T_c)/v0 \approx 6$ \cite{eos_vliegenthartlekkerkerker_jcp2000} for the critical value of the second virial coefficient can serve as an example. Provided that similar estimate for the triple point $T_{tr}$ is given then $T_{c}> T_{tr}$ can give the searched stability criterion for systems where the VL rule can be applied  \cite{liq_vlieglekkvirialzhou_molsimul2007,eos_secondvir_lengmuir2014}.
The very possibility to relate high- and low- density fluid states appears due to so called Zeno-line (ZL) linearity \cite{eos_zenobenamotz_isrchemphysj1990}, which generalizes the Batschinski law for the van der Waals equation \cite{eos_zenobatschinski_annphys1906}. The phenomenological observation states that the line of unit compressibility  
	\begin{equation}\label{eq:z1}
	Z = \frac{P}{\rho\,T} = 1\,,
	\end{equation}
is approximately straight for many fluid systems \cite{eos_zenoline_potentials_jcp2009, eos_zenoapfelbaum1_jpcb2009}, \cite{eos_zenosanchez_jpcb2016,eos_zenolineionic_cpl2017} and can be represented  by the equation: 
\begin{equation}\label{eq:zenovir}
\frac{\rho}{\rho_B} + \frac{T}{T_B} = 1\,.
\end{equation}
Here the corresponding parameters $T_{B}$ and $\rho_{B}$ are calculated using the second $B_2$ and the third $B_3$ virial coefficients \cite{book_hansenmcdonald} only: 
\begin{equation}\label{eq:tbnb}
B_2(T_{B}) = 0\,,\quad \rho_{B}= \frac{T_B}{B_3\left(\,T_B\,\right)}\,\left. \frac{dB_2}{dT}\right|_{T= T_B}\,.
\end{equation}
For example, the generalized thermodynamic similarity reasoning based on \eqref{eq:zenovir} allows to predict high-temperature critical points of some metals \cite{eos_zenoapf_morseiron_jcp2011,eos_apfelberill_jpcb2012,eos_zenoapfelbaumGa_jml2018}. So it is natural to use the ZL regularity \eqref{eq:zenovir} for searching the markers for stability loss of the liquid phase and the corresponding deformation of the gas-liquid binodal. Here another general empiric relation  comes to the scene too. The so called Law of Rectilinear Diameter (LRD) \cite{crit_diam1}:
	\begin{equation}
		\rho_{d} = \frac{\rho_{l}+\rho_{g}}{2\,\rho_c} = 1+A\,
		\left(\, 1-T/T_c \,\right)\, .
		\label{eq:rdl}
	\end{equation}
here $\rho_{l, g}$ is the number density of the liquid and gas phases correspondingly, $T_c, \rho_{c}$ is the
critical temperature and density correspondingly. This relation is valid for many real and model system \cite{eos_zenosanchez_jpcb2016,eos_zenolinereply2sanchez_jpchb2017} outside the fluctuational region \cite{book_patpokr}. 

The approach based on the idea of correspondence between thermodynamic states of fluid and lattice gas where  regularity \eqref{eq:rdl} plays important role was proposed in \cite{eos_zenome0_jphyschemb2010}. We call it fluid - liquid gas global isomorphism. It is based on the hypothesis that the LRD \eqref{eq:rdl} in continuous fluid corresponds to particle-hole symmetry of the  lattice gas model with the Hamiltonian:
\begin{equation}\label{eq:ham_latticegas}
H = -\,\sum\limits_{\left\langle\, ij \,\right\rangle} \,J_{i,j} n_{i}\,n_{j} - h\,\,\sum\limits_{i}\,n_{i}\,,\qquad n_{i} = 0,1
\end{equation}
where $J$ is the energy of attraction between the nearest sites, $n_{i} = 0,1$ and $h$ is the chemical potential of lattice gas. Such correspondence can be expressed as the transformation of state variables. Using the  LRD as the approximation and the concept of the triangle of the liquid-gas state \cite{eos_zenotriapfelbaum_jpchemb2006} one can construct the projective transformation of the lattice gas states $(x, t)$  and
the fluid states $(\rho, T)$:
\begin{equation}\label{eq:projtransfr}
\rho =\, \rho_*\,\frac{x}{1+z \,\tilde{t}}\,,\quad
T =\, T_*\,\frac{z\, \tilde{t}}{1+z \,\tilde{t}}\,,\quad z = \frac{T_c}{T_* - T_c}
\end{equation}
where $x=\left\langle n_{i}\right\rangle$, $ \tilde{t}=t/t_{c}$ is the cite occupation probability and the temperature of the LG reduced to its critical value $t_{c}$. The parameters $T_*$ and $\rho_*$ will be defined in Section~\ref{sec:2dBinadal} differ from the commonly used Boyle parameters \eqref{eq:tbnb}. They determine the Zeno-element 
\begin{equation}\label{eq:zenoelmnt}
\frac{\rho}{\rho_*}+\frac{T}{T_*}=1
\end{equation}
which forms the triangle of liquid-gas states \cite{eos_zenotriapfelbaum_jpchemb2006} in our approach.

One can interpret Eq.~\eqref{eq:projtransfr}
in terms of the relation between the corresponding thermodynamic potentials for
the lattice-gas model $\mathfrak{G}(t,h,\mathcal{N})$ and isomorphic fluid $G(T,\mu,V)$:
\begin{equation}\label{eq:grandpot}
\mathfrak{G}(t,h,\mathcal{N}) = \mathcal{N}\,\mathfrak{g}(t,h)\,,\quad
\text{and}\,\quad G(T,\mu,V) = P(T,\mu)\,V\,,    
\end{equation}
Here $\mathcal{N}$ is the
number of sites in a lattice and $P,\mu$ is the pressure and chemical potential of a fluid. It is natural to state the following relation $\mathcal{N} = \rho_{*} \,V$ between the
extensive variables of these ensembles with $\rho_{*}$ as the characteristic of extremely dense fluid state which corresponds to lattice gas state with fully occupied sites. Using standard thermodynamic relations:
\begin{equation}\label{eq:nx_definitions}
\rho =\f{1}{V}\left.\frac{\partial\, G}{\partial\, \mu}\right|_{T}\,,\quad x
=\f{1}{\mathcal{N}} \left.\frac{\partial\, G}{ \partial\,
h}\right|_{t}
\end{equation}
Eq.~\eqref{eq:projtransfr}
can be obtained from relation between the potentials:
\begin{equation}\label{eq:transfm_potentials}
G (\mu,T,V) =  \mathfrak{G}
\left(\,h(\mu,T),t(T), \mathcal{N}
\,\right)\,\Rightarrow P(\mu,T) = \rho_{*}\,\mathfrak{g}
\left(\,h(\mu,T),t(T)\,\right)\,.
\end{equation}
provided that:
\begin{equation}\label{eq:nmut}
  \rho(\mu,T)/\rho_{*} = x(h(\mu,T),t(T))\,\left(\,1-T/T_* \,\right)\,.
\end{equation}
Taking into account that:
\begin{equation}
\left.\frac{\partial\, }{\partial\, \mu}\right|_{T} =
\left.\frac{\partial\, h }{\partial\, \mu}\right|_{T}
\left.\frac{\partial\, }{\partial\, h}\right|_{t}\,,
\end{equation} 
from
\eqref{eq:nx_definitions},\eqref{eq:transfm_potentials} and
\eqref{eq:nmut} we get the following relation between chemical potentials:
\begin{equation}\label{eq:hmu}
  h(\mu,T) = \left(\,1-T/T_* \,\right)
  \left(\,\mu -\mu_{0}(T) \,\right)\,,
\end{equation}
or, in the inverse form:
\begin{equation}\label{eq:muh}
  \mu -\mu_0(T)  = h\,\left(\,1+z\,t \,\right)\,\,.
\end{equation}
We remind that $h=0$ below CP is the coexistence line for the
LG and is mapped onto the saturation curve of the continuum
fluid. Therefore $\mu_0(T)$ coincides with the chemical
potential $\mu_{s}(T)$  along the saturation curve below the
critical point $T<T_c$. In a supercritical region $\mu_0(T)$ is a continuation of the the diameter and in fact is the Widom-Stillinger line of conjugated states \cite{crit_diampartholewidom_jcp1973}.

Such an approach  was successfully applied and tested for simple fluids with the Lennard-Jones interaction potential and some others \cite{eos_zenomegenpcs_jcp2010,eos_vliegerthartlekkerkerkerme_jcp2011}. Provided that the parameters $T_*$ and $\rho_*$ reflect all relevant information on the attractive part of the interaction  potential we may expect that loosing the stability of liquid phase is reflected in breaking  \eqref{eq:projtransfr} if the attraction is too weak. Previous analysis showed\cite{eos_zenoyukawame_jcp2022} that this is indeed the case for 3D hard-core attractive Yukawa fluids (HCAYF), with the interaction potential:
\begin{equation}\label{eq:yukwpot}
\Phi(r) =\begin{cases}
\infty\,, & \text{if}\quad r< \sigma \\
-\frac{\varepsilon}{r/\sigma} \exp\left(\,-\lambda (r/\sigma -1) \,\right)\,, & \text{if} \quad r\ge \sigma\,,
\end{cases}
\end{equation}
Here the stable liquid phase exists if the screening length $1/\lambda$ is not too small \cite{eos_yukawafluid_jcp1994,liq_dijkstra_hsyukw_pre2002,eos_yukawafluid_molph2007}. So this is a good testing system for checking the behavior of $T_{*}$ and $\rho_{*}$ as markers of the liquid phase instability. Further common dimensionless definition for the temperature $T\to T/\varepsilon$ and the density $\rho\to \rho\,\sigma^{2}$ will be used below.
	
This aim of this paper to add to the results of previous works \cite{eos_zenomegenpcs_jcp2010,eos_zenoyukawame_jcp2022} by considering the gas-liquid equilibrium and its stability for 2D HCAYF using the global isomorphism approach. The structure of the paper is as follows. In Section~\ref{sec:2dBinadal}, we map the binodal of the 2D Ising model (Onsager solution) on 2D HCAYF binodal via \eqref{eq:projtransfr} and checking it using available data of \cite{liq_surftensyukawa_cmp2012}. We also discuss the procedure of extrapolation of the fluid binodal to the low-temperature region. The tangent constructions to the liquid branch of the binodal using $T_{B},n_{B}$ and $T_{*},\rho_{*}$ parameters are compared. In Section~\ref{sec:zenoelmntparam}, we demonstrate the difference between the parameters of the transformation \eqref{eq:projtransfr}  and the ZL parameters \eqref{eq:tbnb} with respect to the stability of the gas-liquid equilibrium. The case of screened Sutherland potential is considered and the range of liquid phase stability is determined. In concluding Section we discuss obtained results which point to \eqref{eq:projtransfr} and  $T_{*},\rho_{*}$ as the relevant characteristics of liquid phase instability for the systems under consideration. 
	
\section{2d HCAYF Binodal in Global Isomorphism approach}\label{sec:2dBinadal}
In this section, we demonstrate the important role of the parameters $T_*$ and $\rho_*$ in the transformation \eqref{eq:projtransfr} for constructing the mapping between the binodals of the Yukawa fluid and lattice gas.  A detailed argumentation regarding the use of the Zeno-element (ZE) parameters in connection with the lattice gas Hamiltonian \eqref{eq:ham_latticegas} was given in \cite{eos_zenoyukawame_jcp2022}. In this way, we consider that the tangent to the liquid branch of the binodal intersects the corresponding coordinate axes at the points $T_*$ and $\rho_*$, which are determined by the following equations:
\begin{equation}\label{eq:tbvdwmy}
B^{vdW}_2(T_{*}) = 0\,,\quad  T_{*} =  T^{(vdW)}_{B}  = \frac{a}{b}\,,
\end{equation}
where
\begin{equation} \label{eq:vdw_ab}
a=\,-\pi\,\int\limits_{\sigma}^{+\infty}\Phi_{\text{attr}}(r)\,r\,dr\,.   
\end{equation}
Here $\Phi_{attr}(r)$ is a attractive part of interaction potential, $\Phi(r)$, $\sigma$ - particle's diameter, so $b = \frac{\pi}{2}\,\sigma^{2}$. The parameter $\rho_*$ is determined by the second and third virial coefficients and the temperature $T_{*}$:
\begin{equation}\label{eq:nbvdwmy}
\rho_*= \frac{ T_* }{B_3\left(\,T_*\,\right)}\,\left. \frac{d\,B_2}{dT}\right|_{T= T_*}\,.
\end{equation}
Thus the parameter $\rho_*$ depends on the total interaction as it represents the dense liquid state where the repulsive potential can not be neglected. It can be considered as the density characteristic of an "analytic continuation" of the saturated liquid densities in the normal liquid range, as these states demonstrate linear temperature behavior. This is consistent with the LRD \eqref{eq:rdl} and leads to the tangent construction to the liquid branch of the binodal at $T\to 0$. The binodal itself practically coincides with the isobar in this region \cite{eos_zenosanchez_jpcb2016}. This allows to find the coefficient of volumetric expansion based on mapping \eqref{eq:projtransfr} \cite{eos_zenoapfvorobisobaric_jcpb2011}. 
	
\begin{figure} 
\hfill
\subfloat[]{\scalebox{0.33}{\includegraphics{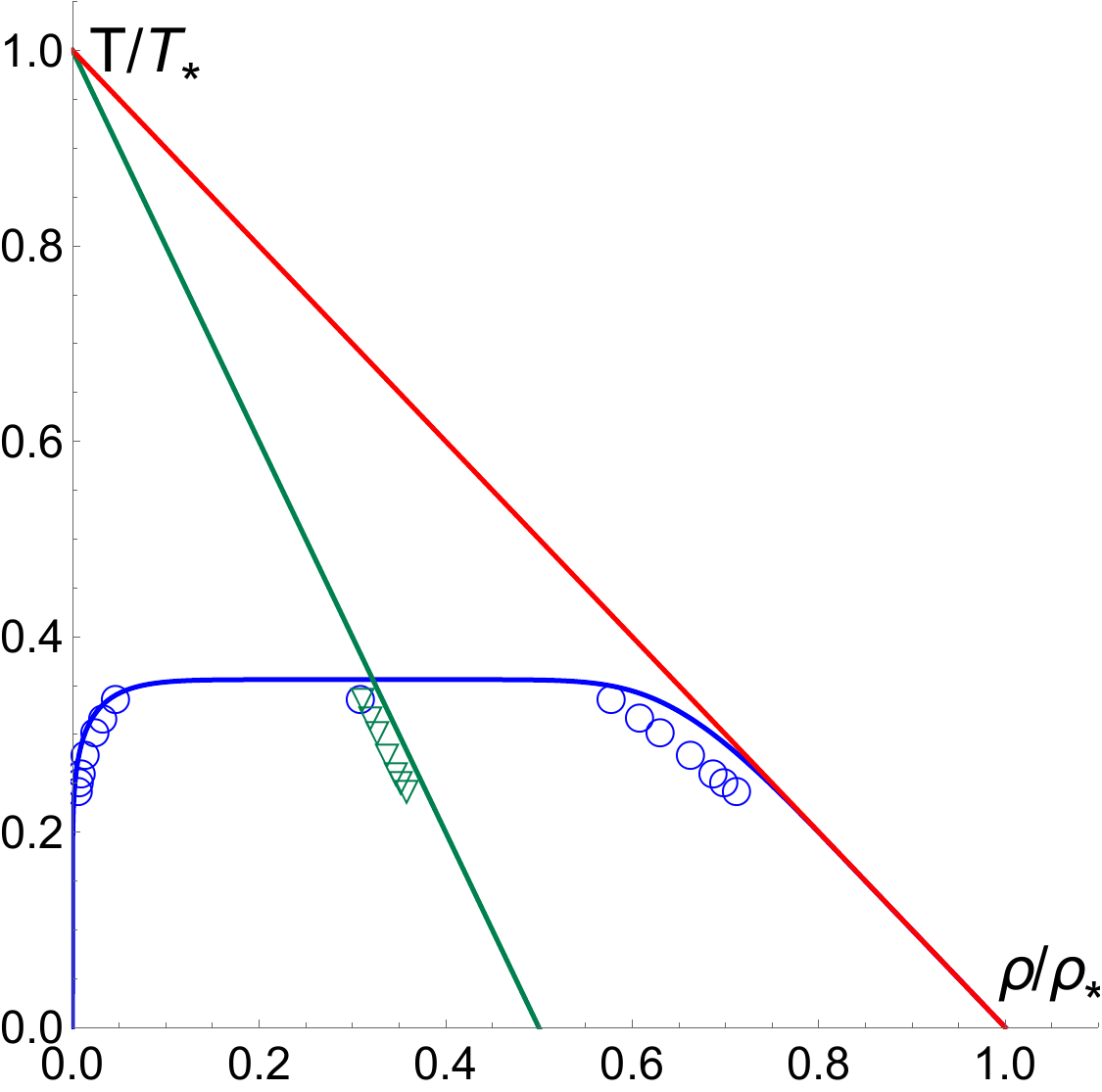}}
\label{fig:onsager_project_1.0}}
\hfill
\subfloat[]{\scalebox{0.33}{\includegraphics{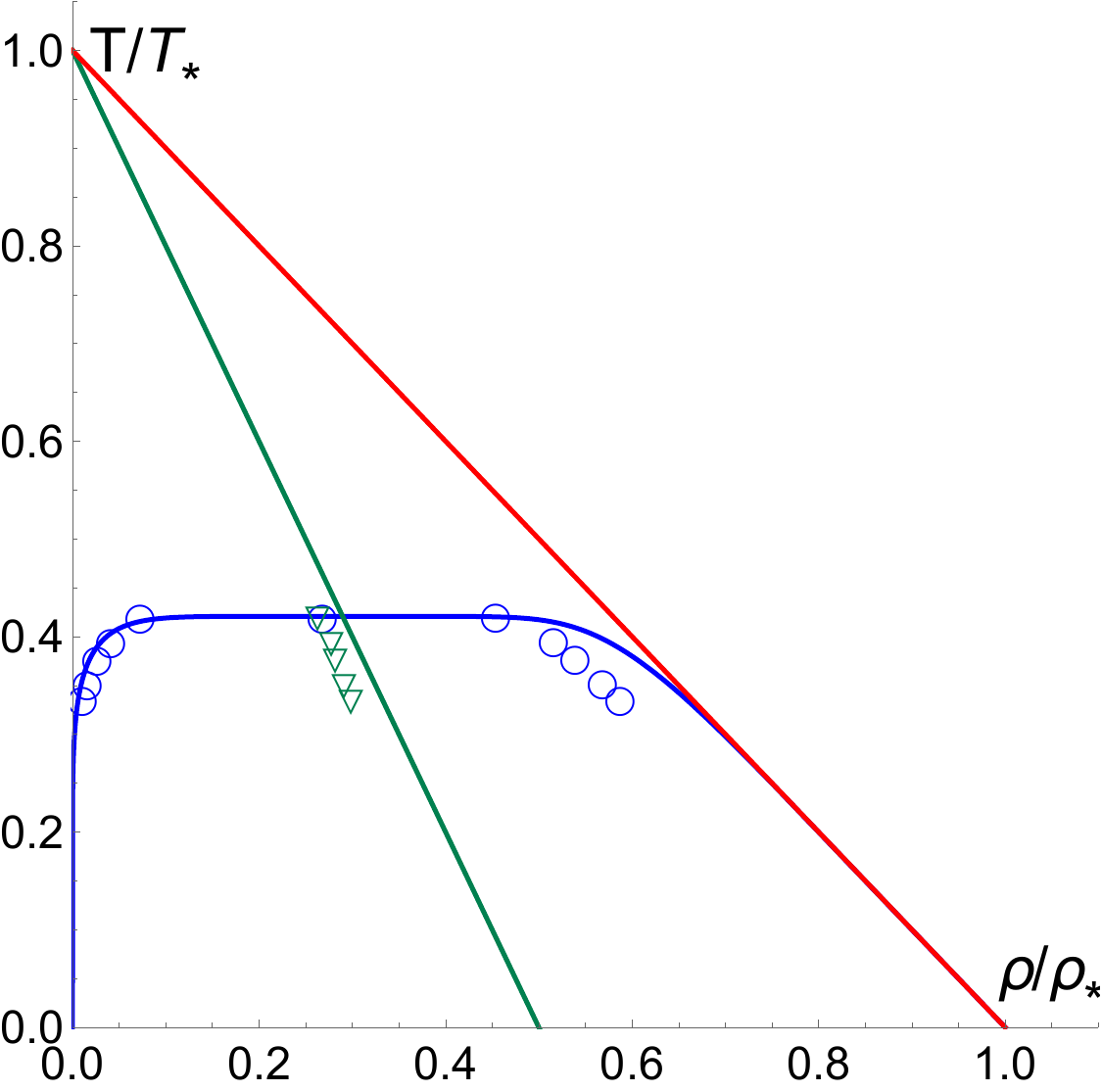}}
\label{fig:onsager_project_1.8}}
\hfill
\caption{Projective transformation applied to Onsager's solution\eqref{eq:onsager}. The empty circles are the data of the numerical simulations\,\cite{liq_surftensyukawa_cmp2012} normalized to the Zeno-element (red) parameters \eqref{eq:tbvdwmy} and \eqref{eq:nbvdwmy}. (a) binodal data for $\lambda=1.0$ and $z\approx 0.51$, (b) binodal data for $\lambda=1.8$ and and $z\approx 0.7$}
\label{fig:onsager_project}
\end{figure}

Based on the equation \eqref{eq:projtransfr} we can check the fluid-lattice gas isomorphism by mapping the binodal of 2D Ising model \cite{crit_onsager_pr1944}:
\begin{equation}\label{eq:onsager}
x=\frac{1}{2}\pm f(t)^{1/8}\,,\quad f(t) = 1 - \frac{1}{\sinh^{4}(2J/t)}
\,.
\end{equation}
onto the binodal data obtained in \cite{liq_surftensyukawa_cmp2012} (see Fig.~\ref{fig:onsager_project}). Note that correct binodal for liquid-gas equilibrium in $2D$ case is very
hard to obtain within some perturbative approach because the non-classical 2D Ising model critical exponent $\beta = 1/8$ makes the binodal dome rather flat (see e.g. \cite{eos_2dlj_canjphys1986,crit_yukawa2d_jcp2018}). That is why some non perturbative approach is needed and global isomorphism provides such a path via \eqref{eq:projtransfr}.
We use the gas branch of the binodal to find the best fit with respect to the parameter $z$ for Eq.~\eqref{eq:projtransfr}. Such a  choice is motivated by the fact that this part of the binodal is more accurately reproduced in numerical simulations. The best result was achieved at $z=0.51$ for $\lambda=1$ (Fig.~\ref{fig:onsager_project_1.0}) and at $z=0.70$ for $\lambda=1.8$ (Fig.~\ref{fig:onsager_project_1.8}).
Now we can estimate locus of critical point for different $\lambda$. From Eq.~\eqref{eq:projtransfr} with taking into account $x_c=1/2\,, \tilde{t}_c = 1$ we get simple formulas for the critical point locus:
\begin{equation}
    \label{eq:tcnc_tsns}
    T_c=\frac{T_*\,z}{1+z}\,\quad \rho_c=\frac{\rho_*}{2(1+z)}
\end{equation}
Our estimate is in Tab.~ \ref{tab:CP}.
 \begin{table}
\centering
    \caption{Comparison of critical parameters.}
    \label{tab:CP}
    \begin{tabular}{|c|c|c|c|c|}
        \hline
        $\lambda$ & $T_c$& $T_c$\,\cite{liq_surftensyukawa_cmp2012} & $\rho_c$ & $\rho_c$\,\cite{liq_surftensyukawa_cmp2012} \\
        \hline
        1 & 0.73 &0.72& 0.39 & 0.36  \\
        \hline
       1.8 & 0.49& 0.5 & 0.44 & 0.42  \\
        \hline
    \end{tabular}
\end{table}

As we can see (see Fig.~\ref{fig:onsager_project}), mapping the Onsager binodal fits well on the numerical data for $\lambda\lesssim 1$.  Also from Fig.~\ref{fig:onsager_project} it follows that the binodal diameter coincides with the median of  liquid-gas triangle with good accuracy, which is consistent with the global isomorphism approach. As the attraction range becomes shorter, the deviation from liquid binodal data increases.
	
In its turn, Eq.~\eqref{eq:nbvdwmy} allows us to construct a binodal tangent to the liquid branch under the condition that $T \to 0$. The recovery of information about the low-temperature part of the liquid branch is based on the idea of the similarity between the states $x \to 0$ and $x \to 1$ in the lattice model \eqref{eq:ham_latticegas}, since the lattice gas is absolutely symmetric to the replacement of the "particle"$\to$" hole". Based on the idea of isomorphism, we can obtain information about the high-density state (liquid branch, $T \to 0$) of the system from the data on the high-temperature and low-density state  $T \to T_*$.
\begin{figure} 
\hfill
\subfloat[]{\scalebox{0.33}{\includegraphics{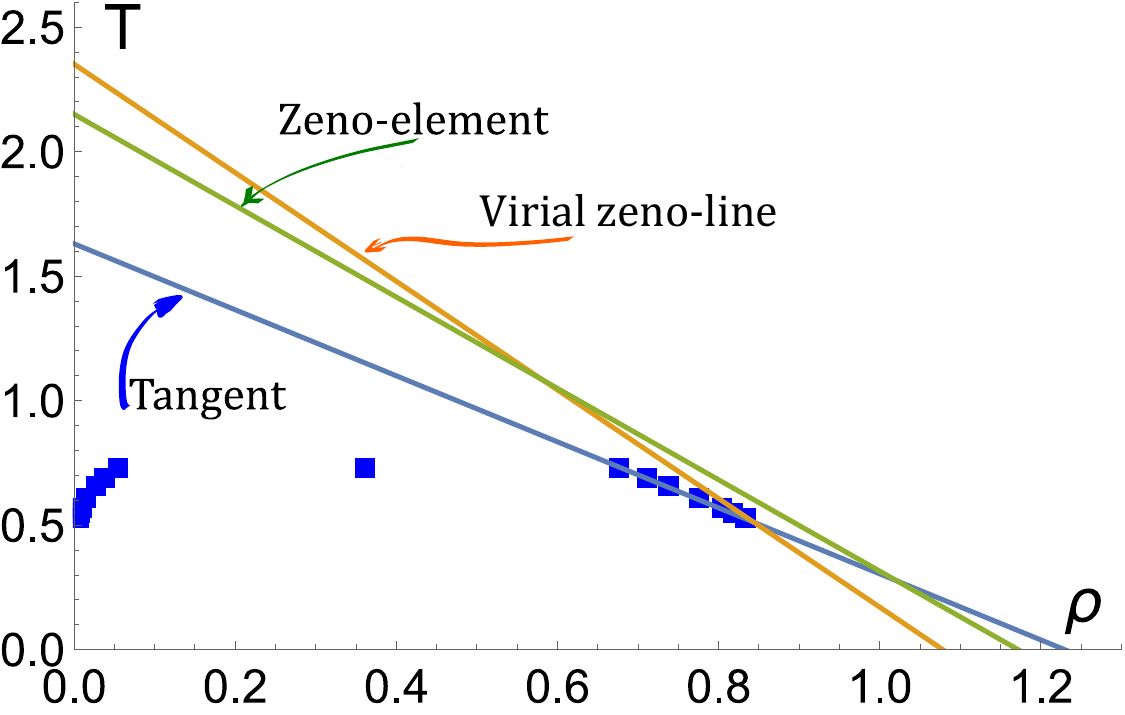}}
\label{fig:num_tan_1.0}}
\hfill
\subfloat[]{\scalebox{0.33}{\includegraphics{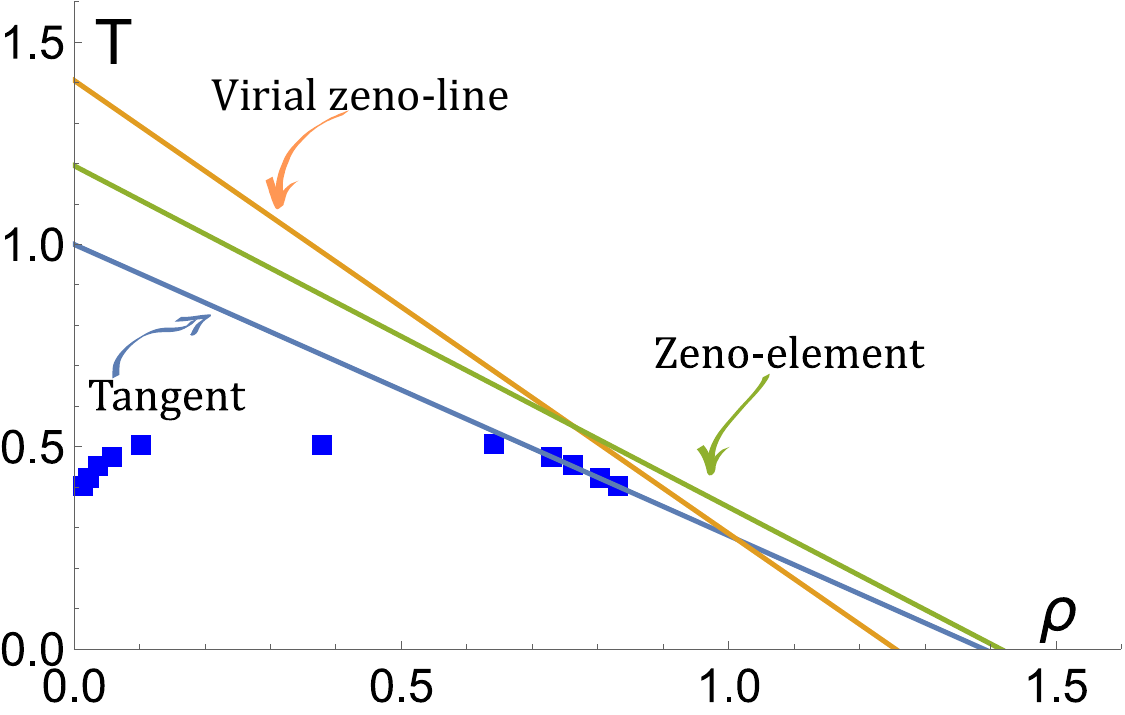}}
\label{fig:num_tan_1.8}}
\hfill
\caption{Comparison of the tangent to the binodal with the Zeno line and the Zeno-element for $\lambda=1.0$ (a) and $\lambda=1.8$ (b), numerical data from the work \cite{liq_surftensyukawa_cmp2012}. The blue line is tangent to the binodal, the green is the Zeno-element, the orange is the Zeno line. 
\label{fig:num_tan}}
\end{figure}

Also we are able to construct a tangent to the liquid branch of the binodal, again verifying our usage $T_*\,,\rho_{*}$ instead of virial Zeno-line parameters $T_B\,,\rho_{B}$. In Fig.~\ref{fig:num_tan} we compare the Zeno-line and the Zeno-element with the tangent constructed by numerical methods. In these graphs, the binodal data are not normalized. We can see that the Zeno-element is located closer to the tangent than the virial Zeno line defined by the classical Boyle parameters \eqref{eq:tbnb}. Note that the tangent in these graphs is built on the basis of numerical data obtained by the Monte Carlo method based on the work mentioned above. They can not be considered  tangent to the binodal at $T \to 0$. As one can see the Zeno-element better describes available data in the context of liquid-gas triangle construction.

We used data for SCAYP to build coexistence curves in this section. From a physical standpoint, it follows the smaller values of $\lambda$ the greater the contribution from the long-ranged attraction to thermodynamic properties. The influence of the soft core repulsion is diminished in such case. On the contrary, the contribution of the soft core increases with the increase of $\lambda$ as demonstrated below (see Fig.~\ref{fig:B3Hard_B3Soft}). Such reasoning is also supported by the results of \cite{crit_yukawa2d_jcp2018} which show the proximity of coexistence curves for $\lambda=1, 1.8$.
\section{Zeno-element parameters of 2d HCAYF}\label{sec:zenoelmntparam}
\begin{figure}[h!]
\hfill
\subfloat[]{\scalebox{0.25}{\includegraphics{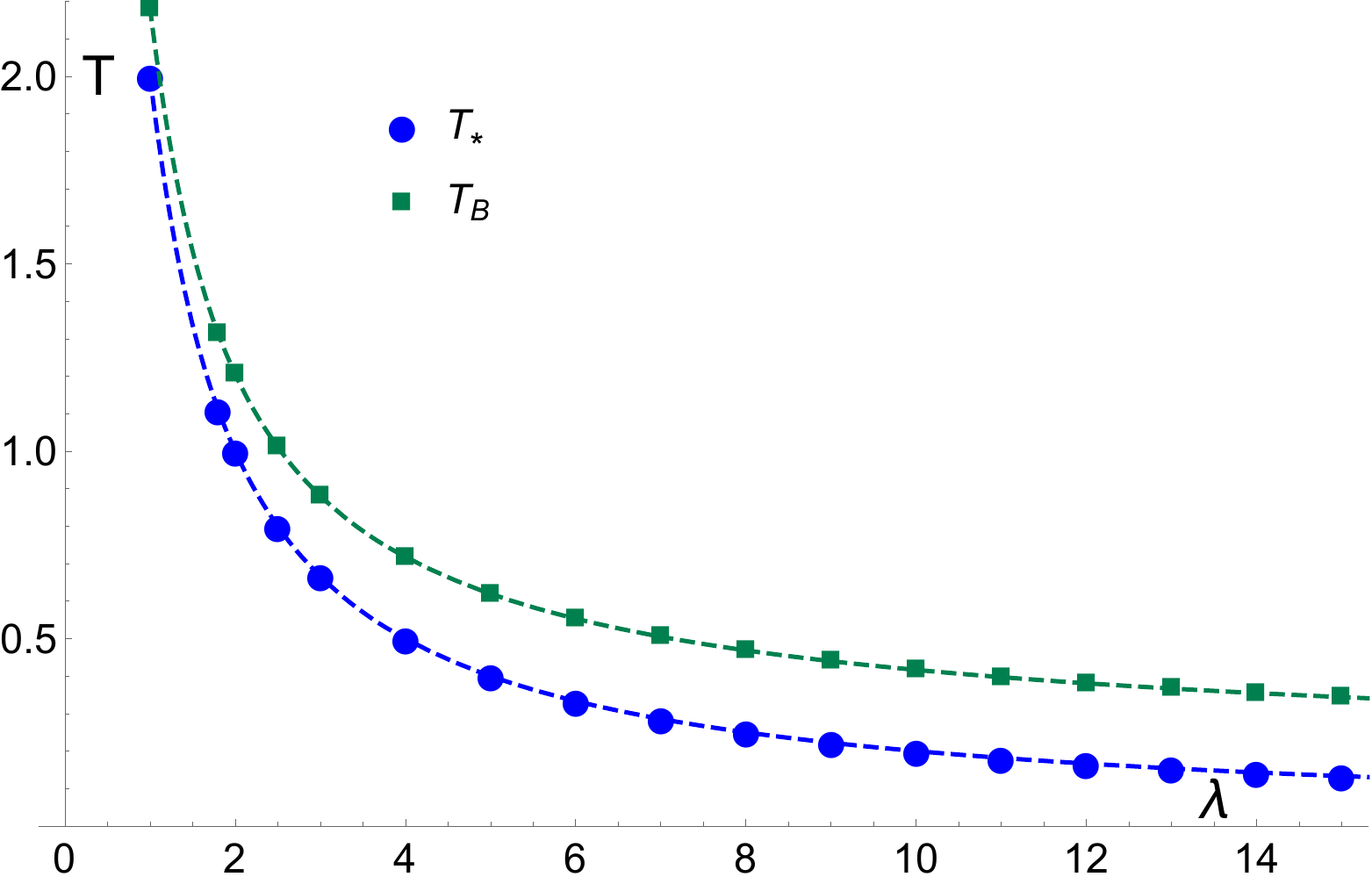}}
\label{fig:t_star_vs_boil}}
\hfill
\subfloat[]{\scalebox{0.25}{\includegraphics{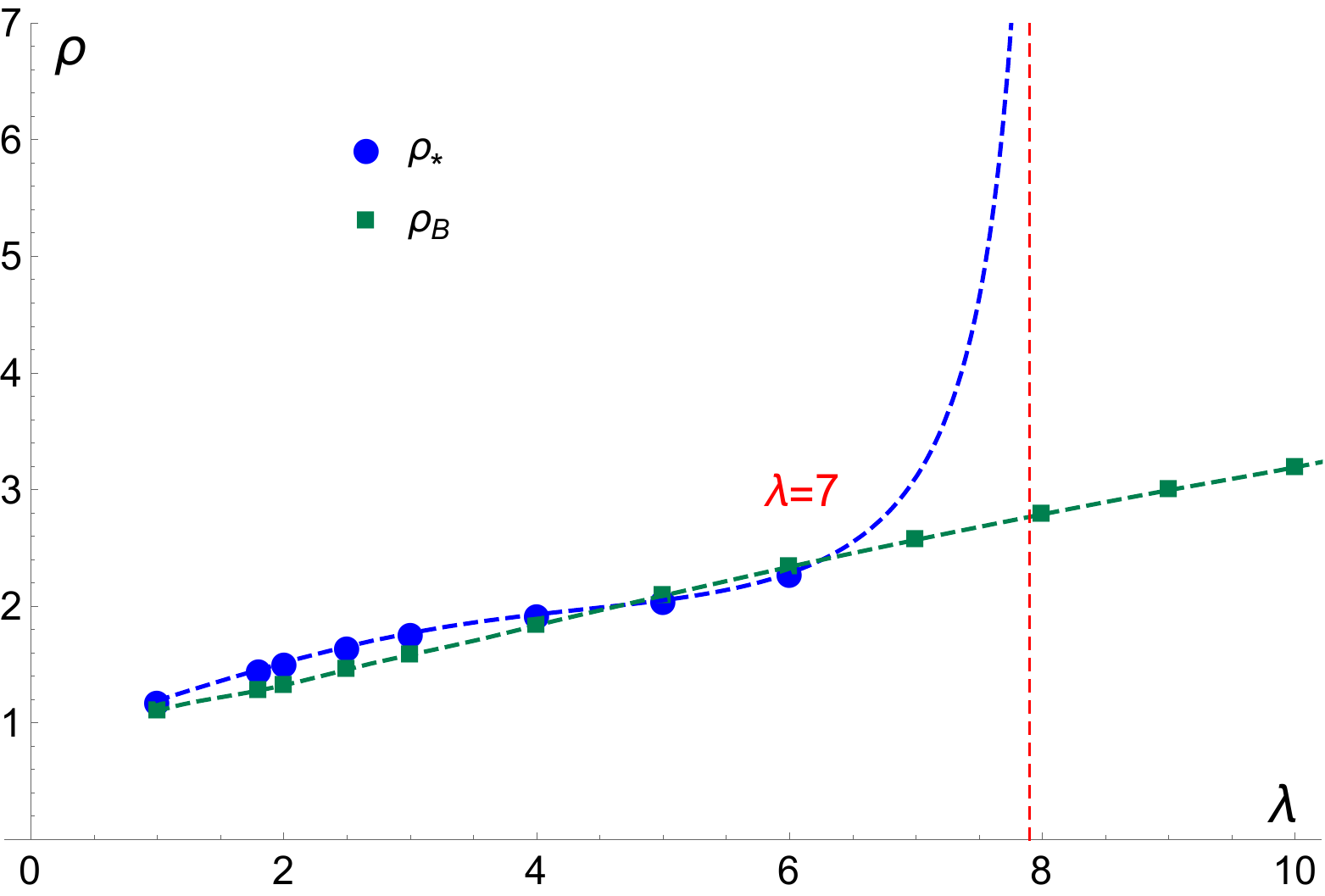}}
\label{fig:rho_star_vs_boil}}
\hfill
\caption{Comparison of the parameters of Zeno-element and Zeno-line; a -  temperature parameters, b - density parameters}
\label{fig:star_vs_boil}
\end{figure}

Here we compare the parameters $\rho_*$ and $T_*$  with the commonly used virial Zeno-line $\rho_B$ and $T_B$ as  functions of the screening parameter $\lambda$. First let us demonstrate that for HCAYF potential the relation $T_{*}<T_{B}$ holds. Clearly, for the HCAYF \eqref{eq:yukwpot} and other models where interaction potential is the sum of the hard core and some smooth potential $\Phi(r)<0,\,\Phi(r\to +\infty)\,\to\,0$ 
\begin{equation}
B_2(T) = \frac{\pi}{2}\,\sigma^2 - \pi\int\limits^{+\infty}_{\sigma}\left(e^{-\Phi(r)/T}-1\right)\,r\,dr
\label{eq:b20}
\end{equation}
Because the integral term in \eqref{eq:b20} is positive  ($\Phi(r)<0, r>\sigma$) and  $x>e^{-x}-1$ if $x>0$, we can conclude that $T_{*}<T_{B}$. The result of direct calculation in Fig.~\ref{fig:t_star_vs_boil} demonstrate this inequality in our case of interest. 
	
The comparison of $\lambda$-dependencies of density parameters $\rho_{*}$ and $\rho_{B}$ shows even more crucial difference between them related to liquid branch instability.  According to \eqref{eq:nbvdwmy} $\rho_{*}$ is determined by the value of $B_3(T_*)$. The situation where $B_3(T)\to 0 $ and $\rho_* \to \infty$ corresponds to the breaking of liquid-gas triangle construction and the mapping \eqref{eq:projtransfr}.  Results of our calculations (see Fig.~\ref{fig:rho_star_vs_boil}) have indicated that the liquid branch of the binodal is unstable if  $\lambda \gtrsim 8$. This result is reasonably consistent  with the data of \cite{liq_surftensyukawa_cmp2012}, where the instability observed to occur at values $\lambda \gtrsim 4$ . 
		
\begin{figure}[h!]
\begin{center}
\includegraphics[width=0.8\linewidth]{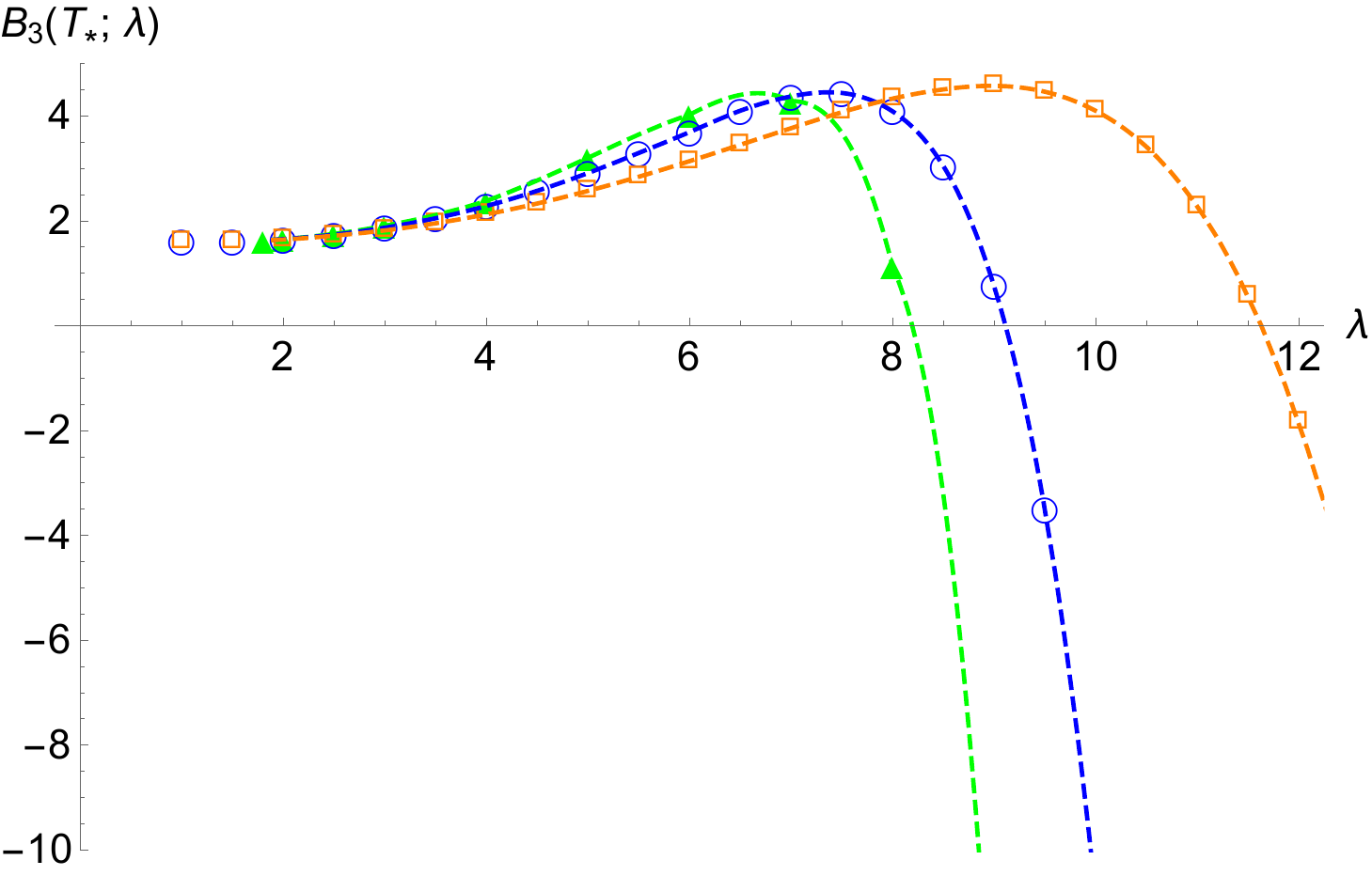}
\caption{Comparison function $B_3(T_*;\lambda)$ for Yukawa with hard core \eqref{eq:yukwpot} and Soft Yukawa potential \cite{liq_surftensyukawa_cmp2012}. Full triangles - Hard Core, empty circles - soft repulsive with exponent $n=255$, empty squares - soft repulsive with exponent $n=100$ }
\label{fig:B3Hard_B3Soft}
\end{center}
\end{figure}

We have also analyzed the dependence of $\lambda_{*}$ that determines the edge of liquid phase stability:
\begin{equation}\label{lambda_crit}
B_3(T_*(\lambda_{*}))=0 \Longrightarrow \rho_* \to \infty
\end{equation}
on repulsion exponent $m$ ($\sim r^{-m}, \,m >100$) which is often used to soften the repulsion in the Yukawa potential. According to our calculations greater repulsion exponent shifts liquid instability to higher values of $\lambda$ i.e. indeed, shorter attraction range (see Fig.~\ref{fig:B3Hard_B3Soft}).
	
\begin{figure}[h!]
\hfill
\subfloat[]{\scalebox{0.25}{\includegraphics{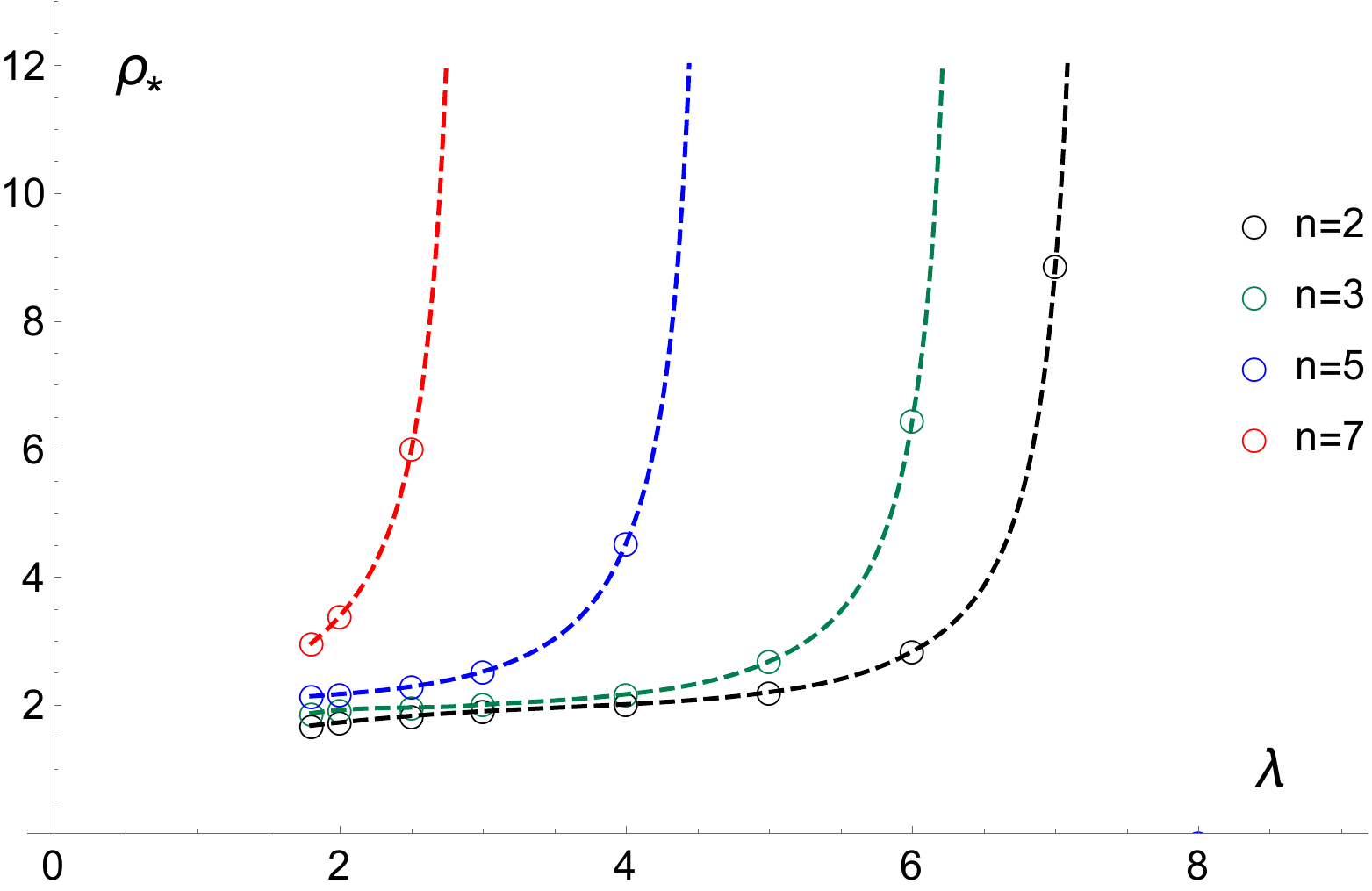}}
\label{fig:rho_star_vs_n}}
\hfill
\subfloat[]{\scalebox{0.25}{\includegraphics{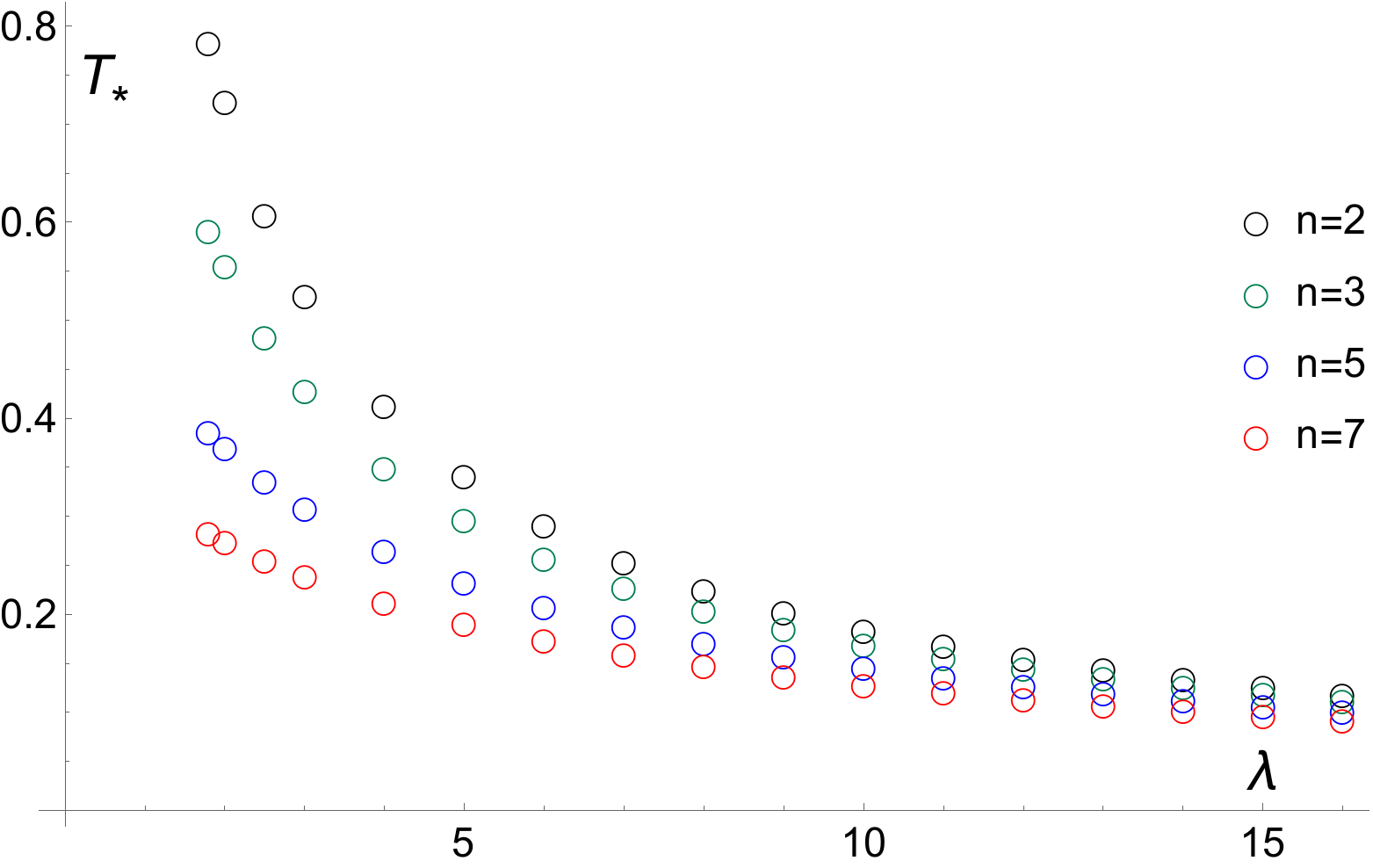}}
\label{fig:t_star_vs_n}}
\hfill
\caption{Parameters of the Zeno-element for the generalized Yukawa potential \eqref{eq:yukwpot_vs_n}; a - density parameters, b - temperature parameters }
\label{fig:star_rho_t_vs_n}
\end{figure}

\begin{figure}[h!]
\hfill
\subfloat[]{\scalebox{0.23}{\includegraphics{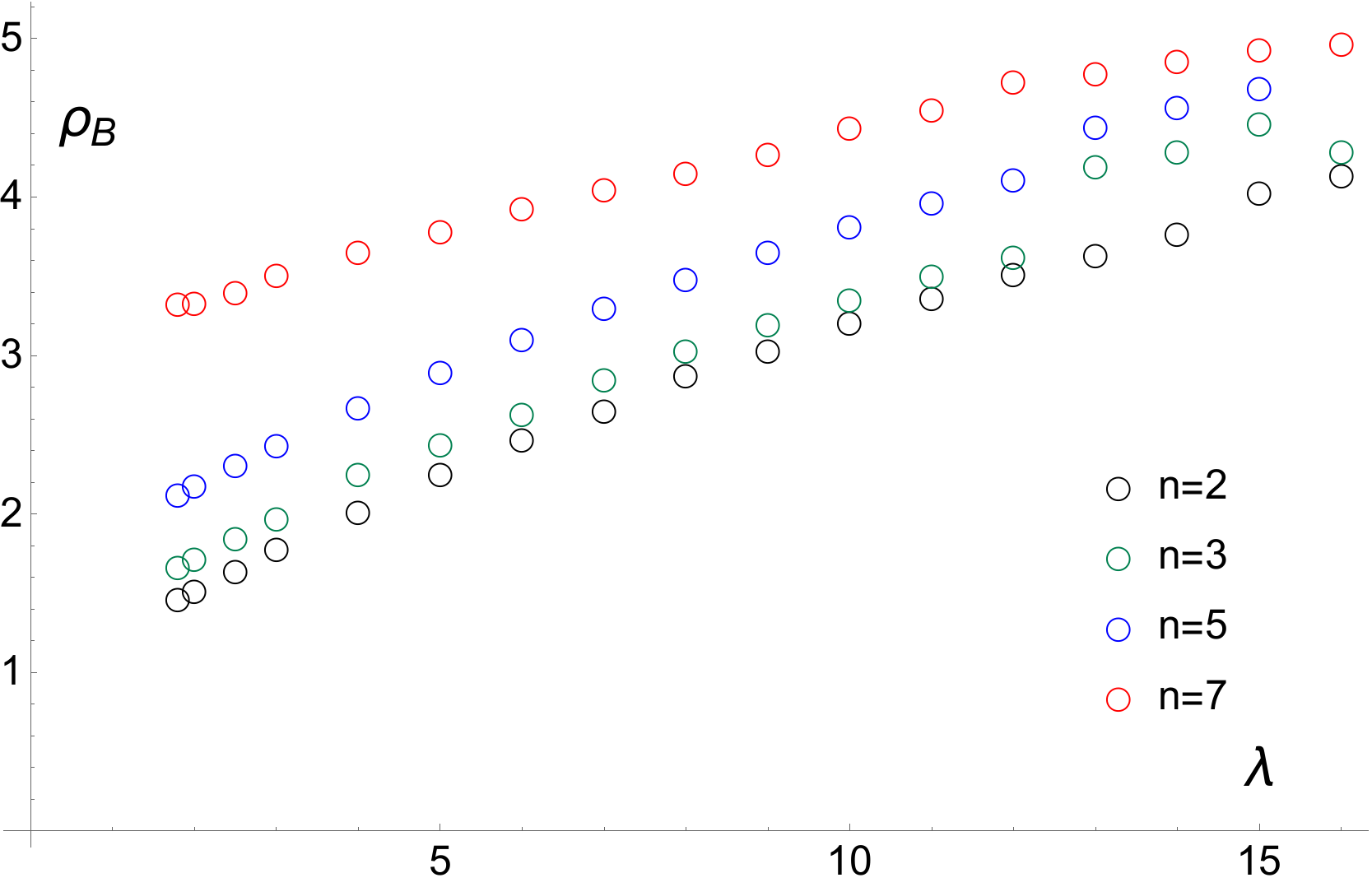}}
\label{fig:rho_boil_vs_n}}
\hfill
\subfloat[]{\scalebox{0.23}{\includegraphics{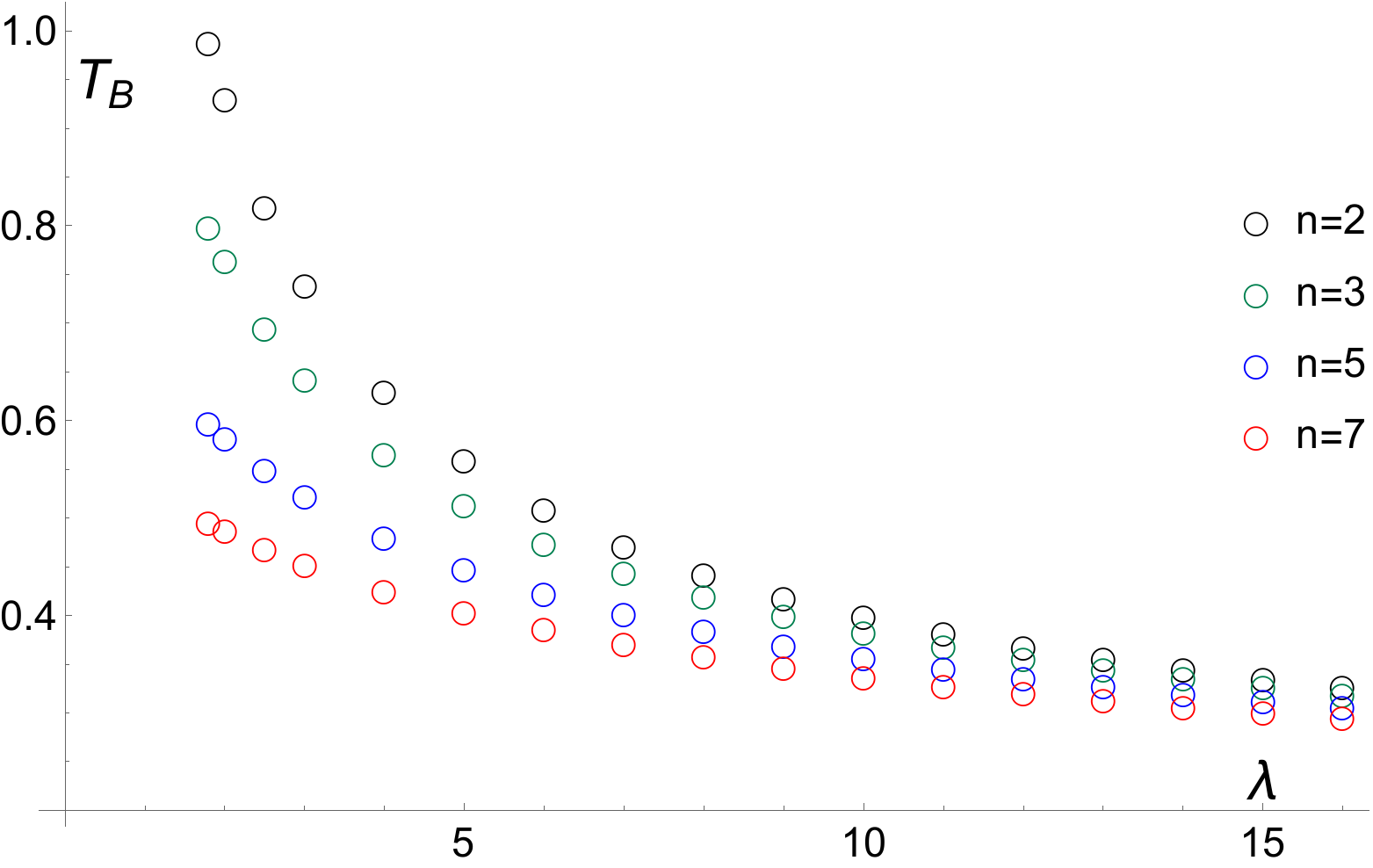}}
\label{fig:t_boil_vs_n}}
\hfill
\caption{Boyle parameters for the generalized Yukawa potential \eqref{eq:yukwpot_vs_n}}
\label{fig:boil_rho_t_vs_n}
\end{figure}

To verify the approach, we  consider the generalized Yukawa potential:
\begin{equation}\label{eq:yukwpot_vs_n}
\Phi(r;\lambda, n ) =\begin{cases}
\infty\,, & \text{if}\quad r< \sigma \\
-\frac{\varepsilon}{(r/\sigma)^{n}} \exp\left(\,-\lambda (r/\sigma -1) \,\right)\,, & \text{if} \quad r\ge \sigma\,,
\end{cases}
\end{equation}
Obviously as $\lambda\to 0$ \eqref{eq:yukwpot_vs_n} turns into the Sutherland potential, for which the stability range of liquid phase is known in 3D $n^*\geq 8$  \cite{eos_longshortcamp_pre2003}. We can not find the data for such potential in 2D but from physical point of view one may expect that the limiting exponent for the liquid phase instability should be less than the 3D value. From physical reasoning one can expect that  if the exponent $n$ increases, the  value $\lambda_{*}$ decreases. Indeed the increase of exponent $n$ in \eqref{eq:yukwpot_vs_n} corresponds to a shorter range of effective attraction. This leads to the fact that a stable liquid phase can exist at smaller values of $\lambda$. Verification of this statement is carried out by calculating the parameters $T_*\,,\rho_{*}$ of the Zeno-element for different values of the exponent $n$ and comparing them with $\rho_{B}$ and $T_B$.  The results of these calculations are presented in Fig.~\ref{fig:star_rho_t_vs_n}, which indicate the validity of using the parameters of the Zeno-element as a marker of the liquid phase instability. Fig.~\ref{fig:boil_rho_t_vs_n} shows the density and temperature of Boil for different values of the $n$ indicator. 
Again  $n_{B}, T_{B}$ do not show any peculiarities and  therefore do not provide any clue to the liquid phase instability. Note that the general relation $T_* < T_B$ holds for the generalized potential for all values of $n$. 

\begin{figure}[h!]
\centering
\includegraphics[width=0.7\linewidth]{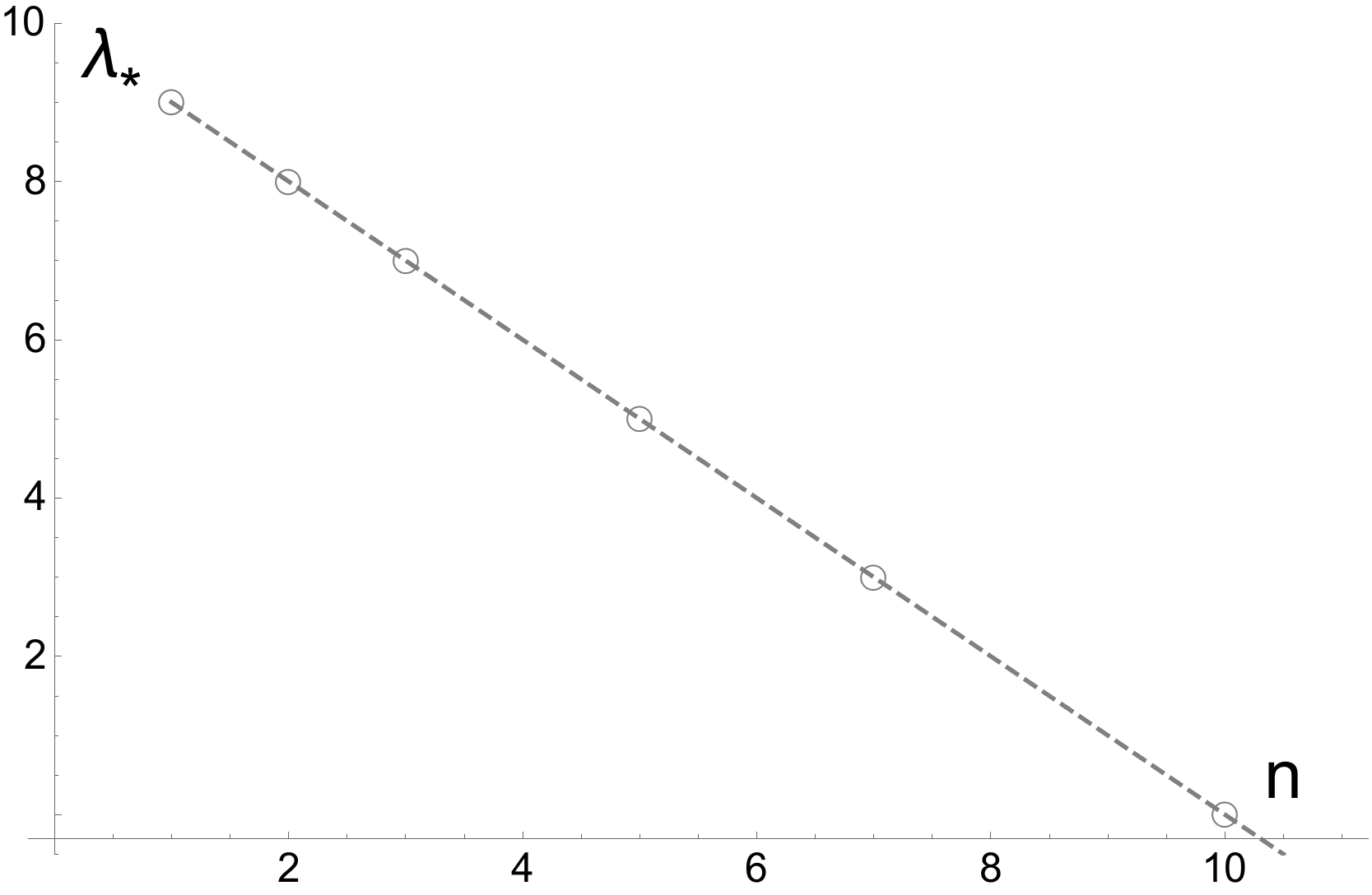}
\caption{Dependence $\lambda_{*}(n)$ for the potential~\eqref{eq:yukwpot_vs_n}, dotted line - interpolation.}
\label{fig:LambdaX}
\end{figure}

The dependence of  $\lambda_{*}(n)$ is demonstrated in Fig.~\ref{fig:LambdaX}. Note that if $\lambda = 0$ i.e. for Sutherland potential the stable liquid branch of the binodal does not exist at $n \gtrsim 10$ in our 2D case which is less than the corresponding exponent in 3D case \cite{eos_zenoyukawame_jcp2022}, as it should be.

\section{Conclusions}
In this work we demonstrate how the data of numerical experiments on the liquid-vapor equilibrium curve of HCAYF fluid in $2D$ can be mapped onto the binodal of the 2D Ising model with the help of projective transformation \eqref{eq:projtransfr}. We have shown that parameters of the Zeno-element $T_{*}$ and $\rho_{*}$ can be used to construct the tangent to the liquid branch of the binodal in low-pressure region (equivalently $T\to 0$). Simple  geometrical picture based on the triangle of liquid-gas states and the LRD \eqref{eq:rdl} make it possible to connect the instability of the liquid phase and the behavior of these ZE-parameters. We attribute the fact that  $\rho_{*}(\lambda_{*}) \to \infty$ so that the binodal tangent at $T\to 0$  becomes horizontal and the coexistence curve does not have a stable liquid branch. Although we use 2D Ising model as obvious candidate for isomorphic lattice model it should be pointed out that the exact correspondence between a fluid and lattice model is not fully understood from the mathematical point of view. There can be other variants of Ising-like model with interactions next to nearest neighbors and different lattice structure. We plan to study these questions further.
	
Obviously, our results can be applied to the study of liquid-gas transition of absorbed monolayers of HCAYF. Analogous studies of Lennard-Jones fluid films on graphite substrates are well known (see e.g. \cite{eos_2dargonadsrb_jpc1951,eos_2dadsorbtion_surfrep1981,eos_2dargonheatcap_prl1984}). It is easy to estimate the bulk-monolayer fluid critical temperature ratio for Lennard-Jones fluid within the global isomorphism approach \cite{eos_zenomegenpcs_jcp2010}:
$$T_{3c}/T_{2c} = 8/3\approx 2.67 $$
which is in rather good agreement with the experimental and simulation data for argon ($T_{3c}\approx 150.8\,K$) $T_{2c}\approx 58 \sim 60\,K$ , xenon ($T_{3c}\approx 289.7\,K$) $T_{2c}\approx 108 \sim 126\,K$ and methane 
($T_{3c}\approx 190.6\,K$) $T_{2c}\approx 69 \,K$
\cite{eos_2dargonmillot_jfr1979,eos_2dxexe_molphys1986,eos_2dargongraphit_jkorean2006,eos_2darxe_condmat2012,eos_2dabsrbch4_prb1986}.
The HCAY potential lacks simple scaling homogeneity property and the determination $z$ becomes more complicated in comparison with the case of Mie-potentials \cite{eos_vliegerthartlekkerkerkerme_jcp2011}. This problem needs separate discussion and we plan to study it in separate publication. Here we provide estimate for this ratio based on our results for $z$ obtained above and numerical simulation from \cite{eos_yukawacorrstates_jcp2008}:
$$
\lambda=1:\, T_{3c}/T_{2c} \approx 3.4\,,\quad 
\lambda=1.8: \, T_{3c}/T_{2c} \approx 2.4
$$
This can be tested either by direct numerical simulation or experimentally with via creation of monolayer of fluid where interparticle interaction can be approximated by the HCAY potential.
\section{Acknowledgement}
This work was completed due to individual (V.K.) Fulbright Research Grant (IIE ID: PS00245791).
%
%

%
\end{document}